\documentclass[preprint,12pt]{elsarticle}

\usepackage{graphicx}
\usepackage{dcolumn}
\usepackage{bm}
\usepackage{color}
\usepackage{booktabs}
\usepackage{amssymb}
\usepackage{amsmath}

\begin{document}

\begin{frontmatter}

\title{Investigation of the magnetic phase transition and magnetocaloric properties of the Mn$_2$FeSbO$_6$ ilmenite}

\author[UU1]{M. Hudl\corref{cor1}}
\ead{matthias.hudl@angstrom.uu.se}
\author[UU1]{R. Mathieu}
\author[UU1]{P. Nordblad}
\author[UU1,Russia1]{S. A. Ivanov}
\author[Russia2]{G. V. Bazuev}
\author[UU2]{P. Lazor}

\address[UU1]{Department of Engineering Sciences, Uppsala University, Box 534, SE-751 21 Uppsala Sweden}
\address[Russia1]{Department of Inorganic Materials, Karpov' Institute of Physical Chemistry, Vorontsovo pole, 10 105064, Moscow K-64, Russia}
\address[Russia2]{Institute of Solid-State Chemistry, Ural Branch of the Russian Academy of Science, 620999 Ekaterinburg, GSP-145, Russia}
\address[UU2]{Department of Earth Sciences, Uppsala University, Villav\"agen 16, SE-752 36 Uppsala Sweden}
\cortext[cor1]{Corresponding author}

\date{\today}

\begin{abstract}
The magnetic phase transition and magnetocaloric properties of both mineral and synthetic melanostibite Mn$_2$FeSbO$_6$ with ilmenite-type structure have been studied. Mn$_2$FeSbO$_6$ orders ferrimagnetically below 270 K and is found to undergo a second-order magnetic phase transition. The associated magnetic entropy change was found to be 1.7 J/kgK for the mineral and 1.8 J/kgK synthetic melanostibite for 5 T field change. For the synthetic Mn$_2$FeSbO$_6$  the adiabatic temperature change was estimated from magnetic- and specific heat measurements and amounts to 0.2 K in 1 T field change. Perspectives of the promising functional properties of Mn$_2$FeSbO$_6$-based materials are discussed.
\end{abstract}


\end{frontmatter}

\section{Introduction}

\begin{figure}[hbt]
		\centering
    \includegraphics[width=0.40\textwidth]{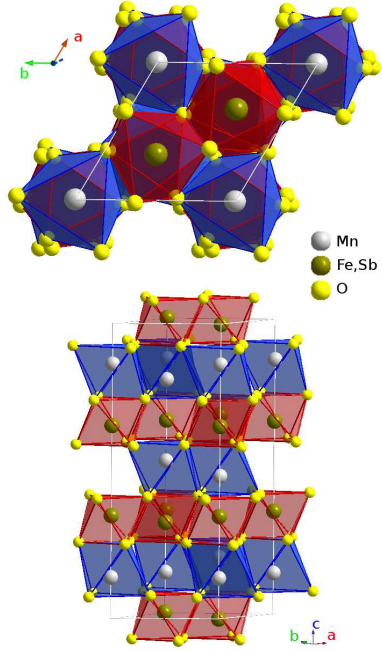}
		\caption{(Color online) Octahedral representation of the melanostibite crystal lattice with ilmenite structure; along $c$-axis (upper panel) and in $ab$-plane direction (lower panel).}
	\label{MFSO0}
\end{figure}

It has been shown that layered magnetic systems can give rise to interesting functional properties e.g. an enhanced magnetocaloric effect such as in Gd$_5$(Si$_{1-x}$Ge$_x$)$_4$ and (Mn,Fe)$_2$(P,Si) compounds \cite{Pecharsky01,Dung11}. The magnetocaloric effect is associated with the temperature change of a magnetic material due to a change of the magnetization under applied magnetic fields. Magnetocaloric materials are intensively studied because of the potential use for cooling and heating applications. The primary advantage of magnetocaloric materials, as opposed to conventional vapor-cycle based technology, is the two orders of magnitude larger entropy density in a solid near its transition temperature compared to a gaseous medium and an efficiency of the magnetic cycle closer to the Carnot limit \cite{Oesterreicher84}. Nevertheless, to use that potential in an environmentally friendly and energy efficient way, materials not based on expensive, toxic and rare elements are needed.\\
\indent Melanostibite is a mineral found in Sj\"ogruvan (\"Orebro, Sweden) and was first described by Lars Johan Igelstr\"om \cite{Igelstrom1892} in 1892. The name melanostibite reflects on the color (\textit{melanos} = black in greek) and composition (\textit{stibium} = antimony in latin) of the material. Melanostibite is a Mn-Sb-oxide with chemical formula Mn$_2$FeSbO$_{6}$ consisting of abundant elements and has a hexagonal crystal structure with space group $R\bar{3}$, which is an isostructural to the ilmenite-pyrofanite group~\cite{Moore68}. The octahedral representation of ilmenite-type structure is shown in Figure \ref{MFSO0}. In this structure Mn and Fe/Sb are ordered in layers along the $c$-axis, whereas Fe and Sb are completely disordered within a Fe/Sb layer. Recently, it has been shown that Mn$_2$FeSbO$_6$ mineral exhibits a ferrimagnetic ordering near 270 K brought forth by the Mn$^{2+}$ and Fe$^{3+}$ ions in the structure~\cite{MFSO-APL}. The magnetic transition in Mn$_2$FeSbO$_6$ is evident from magnetization measurements in low magnetic fields. Nevertheless, its order and the magnetic properties in higher magnetic fields have not been studied yet. A magnetic phase transition close to room temperature and the layered magnetic structure suggesting that this material could give rise to interesting magnetocaloric properties near room temperature. Magnetocaloric oxide materials with perovskite-type structure have been intensively studied~\cite{Rebello08, Rebello11, Sakai09}. To this date there is only a small number of known materials with ilmenite-type structure and a magnetic transition close to room temperature~\cite{Swoboda58}. No study of their magnetocaloric properties have been reported. 

\begin{figure}[htb]
		\centering
    \includegraphics[width=0.843\textwidth]{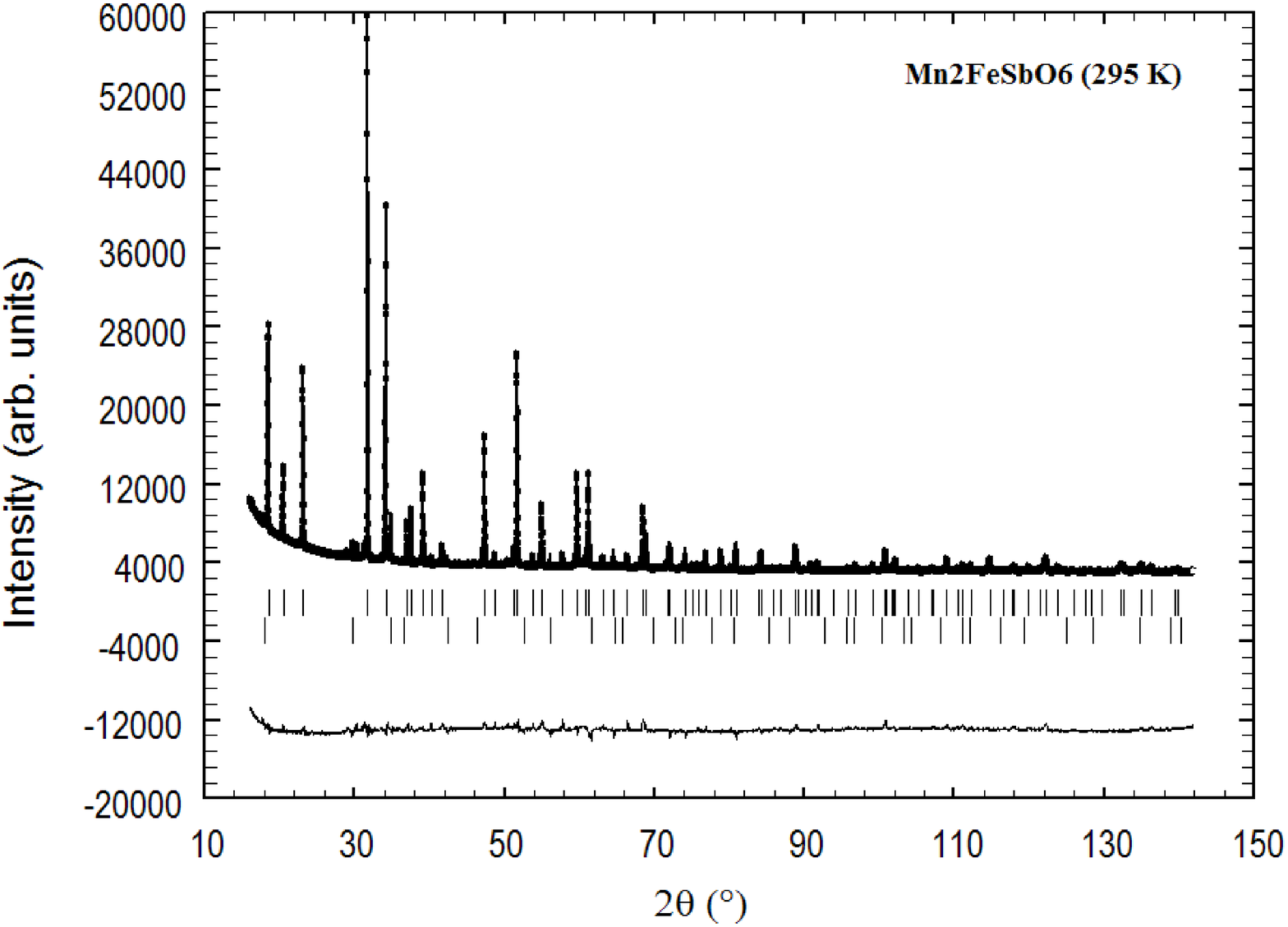}
		\caption{X-ray diffraction pattern and Rietveld refinement (R$_p$=5.64\%, R$_{wp}$=7.32\%, and R$_\beta$=3.18\%) for synthetic melanostibite at 295 K.}
		\label{MFSO1}
\end{figure}

In this article we report on the magnetic phase transition and the magnetocaloric properties of both mineral and synthetic Mn$_2$FeSbO$_6$ with ilmenite-type structure. Our results are based on magnetic- and specific heat measurements. The magnetic entropy changes $\Delta S_M$ for the mineral (synthetic) melanostibite sample was found to be 1.7 (1.8) J/kgK and 0.51 (0.46) J/kgK for field changes of 5 and 1 Tesla, respectively. Although $\Delta S_M$ is relatively small compared to reference magnetocaloric materials such as Gd, we believe that our results indicate that new materials based on ferri-/ferromagnetic ilmenites such as Mn$_2$FeSbO$_6$ with improved magnetocaloric properties near room-temperature could be designed.

\section{Experimental methods}

Synthetic Mn$_2$FeSbO$_6$ samples were fabricated using conventional solid state reaction (including terminatory sintering at temperatures between 1300 and 1350 $^{\circ}$C) and thermobaric treatment under a pressure of 3 GPa at 1000 $^{\circ}$C for 30 min.  Detailed synthesis conditions are reported in Ref. \citet{MFSO-APL} and \citet{Bazuev96}.
The phase composition of the synthetic samples was studied using powder x-ray diffraction (XRD) method on a D8 Bruker diffractometer with CuK$\alpha 1$ radiation. The ICDD PDF4 database of standard powder patterns (ICDD, USA, Release 2009) was used to identify possible impurity phases. The cation stoichiometry of the mineral and synthetic samples was checked using microprobe EDX analysis (average of 20 points). Magnetization measurements for all samples were done using a SQUID magnetometer (MPMS-XL) from Quantum Design Inc (QD). For the magnetization measurements of the natural sample several small ($\approx$10-500 $\mu$m) crystals of the mineral melanostibite were assembled together(total weight of 3.7 mg). Specific heat measurements were performed using a relaxation method on a Physical Properties Measurement System (QD PPMS 6000).

\begin{figure}[htb]
		\centering
    \includegraphics[width=0.543\textwidth]{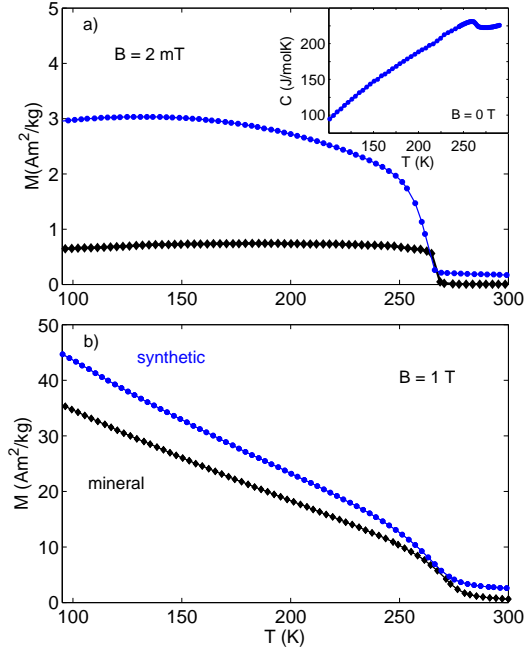}
		\caption{(Color online) Field-cooled magnetization data recorded on cooling (FCC) as a function of temperature for a natural mineral (\textcolor{black}{$\blacklozenge$}) and synthetic (\textcolor{blue}{$\bullet$}) sample melanostibite measured in a) 0.02 mT and b) 1 T. The inset shows the specific heat of the synthetic sample measured in zero magnetic field.}
		\label{MFSO2}
\end{figure}

Incidentally, the melanostibite mineral was found ten years after the discovery of the magnetocaloric effect by E. Warburg in 1881~\cite{Warburg1881}. The magnetocaloric effect is related to the coupling of the spin system with the lattice vibrations of the material and can be represented by the change of magnetic part of the total entropy $\Delta S_{M}$. 
 
For an adiabatic magnetization process the change of the magnetic entropy part causes an identical change in the lattice part of the total entropy - evident as a change in temperature $\Delta T_{ad}$ \cite{Tishin03}.  The magnetic entropy change for a magnetic material undergoing a second-order magnetic phase transition can indeed be derived from Maxwell-relations \cite{deOliveira08}.  Hence, $\Delta S_{M}$ for a magnetic field change from $H_i$ to $H_f$ can be calculated using

\begin{equation}
	\Delta S_{M} = \mu_{0}\int^{H_{f}}_{H_{i}}\left(\frac{\partial M}{\partial T}\right)_{H} dH.
	\label{DeltaSM}
\end{equation}

\begin{figure*}[thb]
			\centering
      \includegraphics[width=0.985\textwidth]{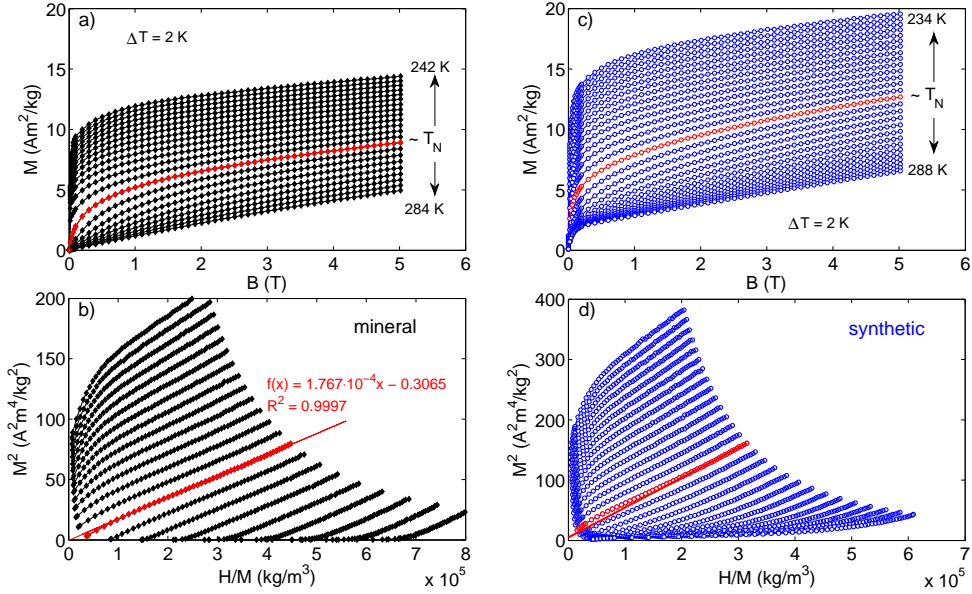}
	\caption{(Color online) Magnetization as a function of magnetic field for a) the natural mineral and c) the synthetic sample melanostibite around $T_N$. Corresponding Arrott plots for b) the natural mineral and d) synthetic melanostibite. A linear fit for the mineral sample near $T_N$ is shown in b),  R$^2$ refers to the accuracy of the least-square fitting.}
	\label{MFSO3}
\end{figure*}

From isothermal magnetization measurements as a function of the magnetic field, $M(H)|_{T}$,  $\Delta S_{M}$ was calculated using the numerical integration 

\begin{equation}
	\Delta S_{M} (T) = \mu_{0}\sum_{i,j}\left(\frac{M_{i+1}-M_{i}}{T_{i+1}-T_{i}}\right) (H_{j+1}-H_{j}),
	\label{DeltaSMnum}
\end{equation}

where $i$ is the number of isothermal measurements, ($T=(T_{i+1}+T_{i})/2$), and $j$ labels the magnetic field increments ($\Delta H = \sum_{j}H_{j}$). In terms of evaluating the nature of a magnetic phase transition, a technique proposed by A. Arrott \cite{Arrott57,Arrott67} can be used. The field dependence of the magnetization can be expressed as a power series ($H = \left(1/\chi\right) M + \beta M^3 + \gamma M^5 + \ldots$) of the order parameter $M$. At the Curie point the magnetic susceptibility approaches infinity and $\left(1/\chi\right) = 0$, which should result in straight lines in the ``critical region" around the phase transition when plotted as $M^2$ vs. $H/M$ (mean field approximation). Analyzing the slope of data plotted as $M^2$ vs. $H/M$ indicates the order of phase transition (Banerjee criterion) \cite{Banerjee64}. A positive slope implies a second-order-continuous phase transition, whereas an S-shaped negative slope suggests a first-order-discontinous phase transition. 

\section{Results and discussion}

An X-ray diffraction pattern for synthetic melanostibite is shown in Figure \ref{MFSO1}. The main phase of the synthesized sample crystallizes in the hexagonal ilmenite-type structure (space group R$\bar{3}$, $a$ = 5.237(1) \AA\ and $c$ = 14.349(2) \AA) similar to the single-crystal mineral of Mn$_2$FeSbO$_6$. From Rietveld refinement,with R-factors R$_p$=5.64\%, R$_{wp}$=7.32\%, and R$_\beta$=3.18\%, a minor amount ($\approx$ 2\%) of a secondary phase of the MnFe$_2$O$_4$ spinel, which is ferrimagnetic below $T_N$=570 K, was estimated. 

In Figure \ref{MFSO2}, the magnetization curves for a mineral (black curve) and a synthetic sample (blue curve) of melanostibite are shown. In small magnetic fields (2 mT) both mineral and synthetic Mn$_2$FeSbO$_6$ show a sharp phase transition. The transition temperature for the mineral is found to be $\sim$268 K and for the synthetic sample slightly lower at $\sim$264 K. The low temperature magnetization for a 5 Tesla field revealed a moment of about 3.7 $\mu_B/f.u.$ for the mineral and 4.35 $\mu_B/f.u.$ (63.1 Am$^2$/kg) for the synthetic melanostibite at 10 K and is in agreement with a ferrimagnetic arrangement of 2 Mn$^{2+}$ and 1 Fe$^{3+}$ cations (all carrying $S=5/2$, 5 $\mu_B$) per formula unit \cite{MFSO-APL}. For a magnetic field of 1 T, Figure \ref{MFSO2}b, the transition appears broader and the magnetization is increasing linearly below $T$$_N$ due to the ferrimagnetic type ordering. The inset in Figure \ref{MFSO2}a shows the specific heat for the synthetic melanostibite in 0 T field.

\begin{figure}[thb]
		\centering
    \includegraphics[width=0.543\textwidth]{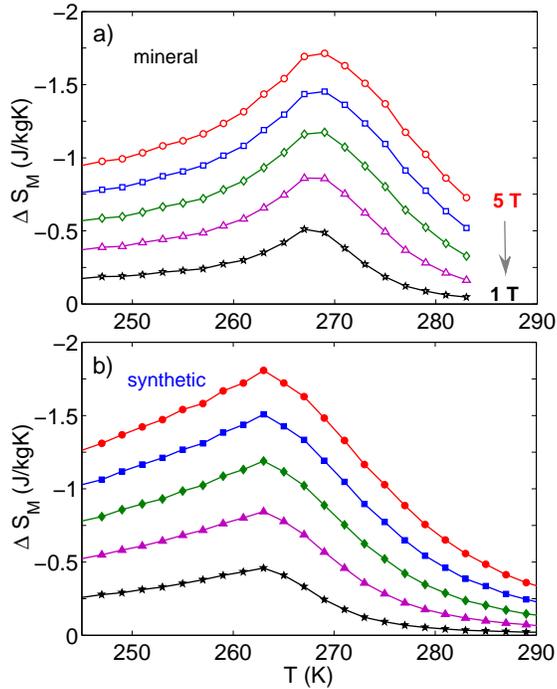}
		\caption{(Color online) Magnetic entropy change $\Delta S_M (T)$ for a) the natural mineral and b) the synthetic sample of melanostibite under a magnetic field change of 1, 2, 3, 4, and 5 T (arrow denotes descending fields). }
		\label{MFSO4}
\end{figure}

From isothermal magnetization measurements around the magnetic phase transition, the Arrott plot can be constructed \cite{Arrott57}. Isothermal magnetization data and Arrott plots for both the mineral and synthetic melanostibite are shown in Figure \ref{MFSO3}. As can be seen in Figure \ref{MFSO3}b and \ref{MFSO3}d a positive slope in the entire data range and therefore a second order nature of the magnetic phase transition is evident for both the melanostibite mineral and synthetic sample (Banerjee criterion). Furthermore, the presented data feature a set of parallel straight lines around the transition temperature which indicate a mean-field-like behavior of the phase transition. For the mineral melanostibite, the behavior of the data is fairly linear as indicated by red line (linear fit) shown in Figure \ref{MFSO3}b. In contrast, some discrepancies presumably due to the presence of the secondary MnFe$_2$O$_4$ phase, were observed for the synthetic sample. The MnFe$_2$O$_4$  phase, as uncovered by our Rietveld analysis, is also evident in the isothermal magnetization data as a hysteresis background at the highest temperature of 288 K in Figure \ref{MFSO3}d. This is considerably above the transition temperature of the melanostibite main phase.   

\begin{figure}[thb]
		\centering
     \includegraphics[width=0.543\textwidth]{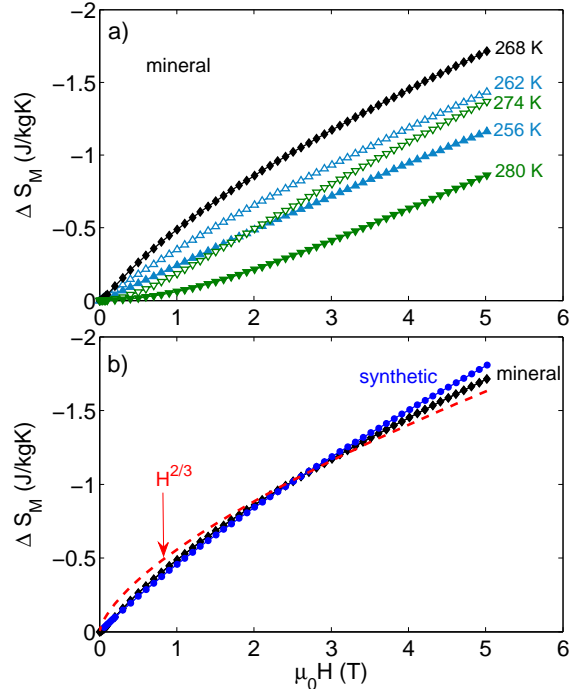}  
		\caption{(Color online) a) Magnetic field dependence of the magnetic entropy change  $\Delta S_M (H)$ for the natural mineral at different temperatures. b)  $\Delta S_M (H)$ for the natural mineral (\textcolor{black}{$\blacklozenge$}) and synthetic (\textcolor{blue}{$\bullet$}) melanostibite at their respective transition temperature and the $H^{2/3}$ power law dependence (dashed lines).}
		\label{MFSO5}
\end{figure} 

The magnetocaloric properties for the mineral are displayed in Figure \ref{MFSO4}a. Melanostibite undergoes a second-order phase transition, hence the peak $\Delta S^{pk}_M$ in the magnetic entropy change as function of temperature occurs at the phase transition temperature and is not shifted with increasing magnetic field changes. Nonetheless, the transition was found to be quite sharp and the magnetic entropy change $\Delta S_M$ is estimated to be 1.7 J/kgK and 0.51 J/kgK for magnetic field changes of 5 T and 1 T,  respectively.  The magnetic entropy change for the synthetic sample can be seen in Figure \ref{MFSO4}b. The magnetic entropy change for the synthetic sample under a magnetic field change of 5 T and 1 T was found to be 1.8 J/kgK and 0.46 J/kgK, respectively. Hence, both samples show comparable values for the magnetic entropy change of the order of 2 J/kgK for a magnetic field change of 5 T. Interestingly, Mn$_2$FeSbO$_6$ crystallizes in a layered structure (c.f. Fig. \ref{MFSO0}) in which the inter-layer coupling of the Mn and Fe/Sb layers is antiferromagnetic, whereas the intra-layer coupling is ferromagnetic. It is well know that the carrier doping and associated chemical pressure may change the inter-layer coupling towards a ferromagnetic behavior, e.g. in perovskite manganites \cite{Okuda99}. Therefore, it is reasonable to expect that cation substitution could be employed to enhance the ferromagnetic interaction in Mn$_2$FeSbO$_6$, yielding improved magnetocaloric properties. 

\begin{figure}[thb]
		\centering
    \includegraphics[width=0.543\textwidth]{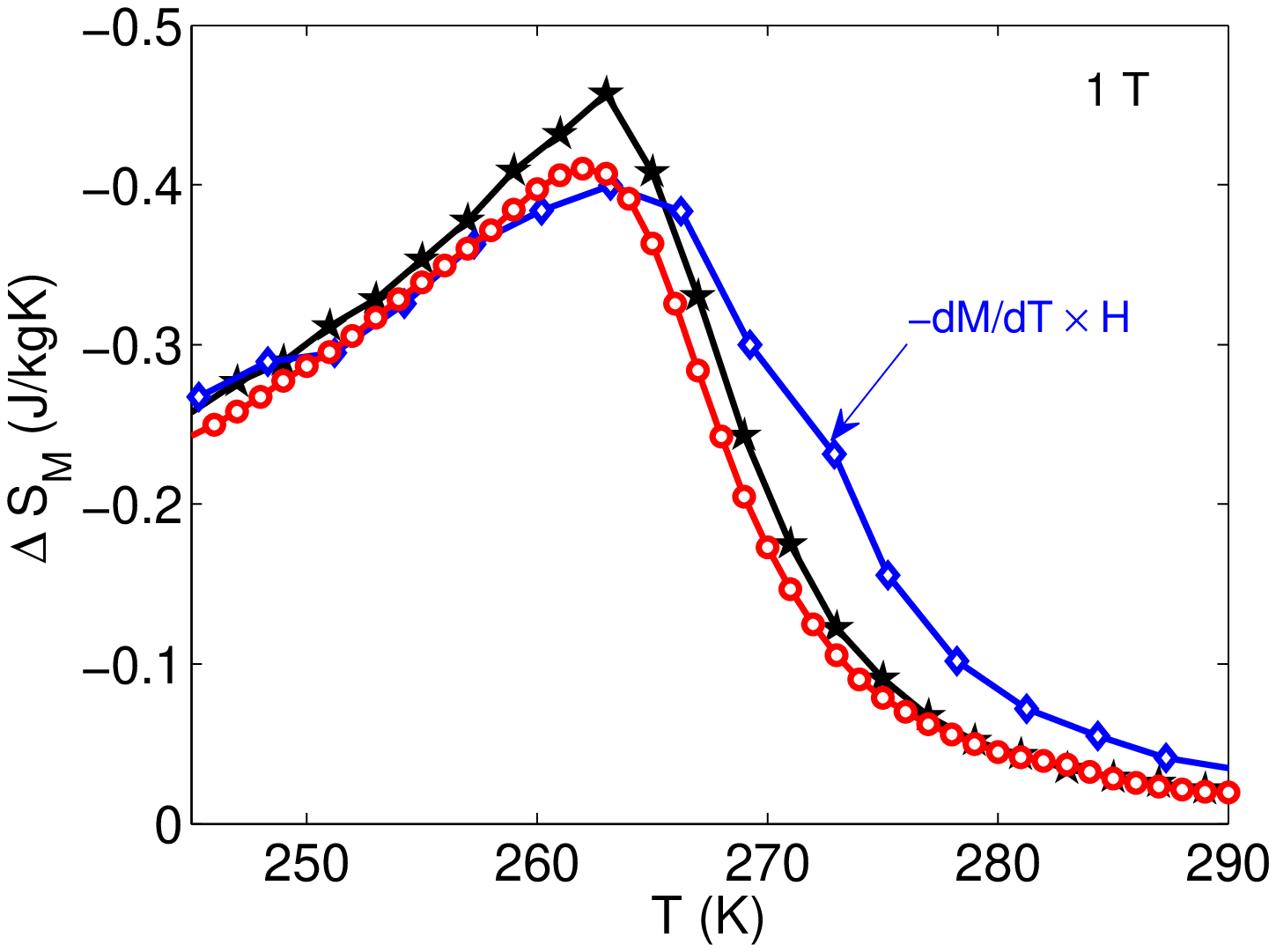}  
		\caption{(Color online) The magnetic entropy change $\Delta S_M$ under 1 T field change as derived from specific heat (\textcolor{red}{$\circ$}) and $M$ vs. $H$ (\textcolor{black}{$\star$}). $-d$$M$/$d$$T$ $\times$ $H$ was derived from $M$ vs. $T$ in 1 T applied field(\textcolor{blue}{$\diamond$}).}
		\label{MFSO6}
\end{figure} 

\begin{table*}[t]
\caption{Magnetic and structural properties for mineral and synthetic Mn$_2$FeSbO$_6$ (MFSO).}
\label{table-Psamples}

\begin{tabular}{|c|c|c|}
\hline 
\bf{Properties} & \bf{ mineral Mn$_2$FeSbO$_6$ } & \bf{ synthetic Mn$_2$FeSbO$_6$ } \\
\hline 
Mn:Fe:Sb & 2.03:0.96:0.99 & 2.05:0.91:1.04 \\
Space group & R$\bar{3}$ & R$\bar{3}$  \\
$a$ (\AA) & 5.226(1)& 5.237(1)\\
$c$ (\AA) & 14.325(2)& 14.349(2)\\
\hline
$T_N$ (K)& 268(2) & 264(2)\\
$\Delta S_{M}^{Pk}$$|_{5 \text{T}}$ (J/kgK) & 1.7 & 1.8 \\
$\Delta S_{M}^{Pk}$$|_{1 \text{T}}$ (J/kgK) & 0.51 & 0.46 \\
$\Delta T$$|_{1 \text{T}}$ (K)& $-$ & 0.2 \\
\hline 
\end{tabular}

\label{comp}
\end{table*}
 
The magnetic field dependence of the magnetocaloric response is shown in Figure \ref{MFSO5}. In agreement with the work of Oesterreicher and Parker \cite{Oesterreicher84}, a power law dependence for the magnetic entropy change $\Delta S \propto a H^n $ is observed. At the transition temperature the exponent $n$ for a system with well localized magnetic moments is expected to be aproximately $n$=2/3. In Figure \ref{MFSO5}b the magnetic field dependence of the magnetic entropy change at the phase transition for the mineral (268 K) and synthetic (264 K) melanostibite as well as the $H^{2/3}$ power law are shown. It is evident that the experimental data at the transition temperature for mineral and synthetic Mn$_2$FeSbO$_6$ follow a $H^{2/3}$ dependence and the magnetic moments are localized.

Figure \ref{MFSO6} depicts the magnetic entropy change $\Delta S_M$ for a 1 T field change derived from measurements of the specific heat and isothermal $M$ vs. $H$ curves. The $\Delta S_M$  curve from specific heat measurements was shifted with a constant offset to match the $\Delta S_M$ at 290 K obtained from $M$ vs. $H$ measurements and Equation \ref{DeltaSMnum}. From a single $M$ vs. $T$  measurement in 1 T applied field $-d$$M$/$d$$T$ $\times$ $H$ was calculated. This quantity is of the same unit as $\Delta S_M$. One can conclude that all three curves reproduce the qualitative behavior for the temperature dependence of the magnetic entropy change and only differ slightly in the absolute value for the $\Delta S_M$. It is interesting to note that a single $M$ vs. $T$  measurement at constant magnetic field gives an acceptable estimate of the magnetic entropy change for a second-order magnetocaloric material. 

From the magnetic entropy change and the specific heat as function of temperature and magnetic field one can derive an estimate for the adiabatic temperature change. The magnetocaloric effect in terms of the change in temperature $\Delta T$ under a 1 T field change for the synthetic melanostibite sample can be estimated to
\begin{equation}
	\Delta T = - \Delta S_{M} \frac{T}{C_{P}(264\text{K, }1\text{T})} \approx 0.2 \text{ (K)},
	\label{DeltaT}
\end{equation}
where $C_{P}(264\text{K, }1\text{T})=231\text{ J/molK}$ is the specific heat measured at 264 K in 1 T applied field. An overview of the magnetocaloric properties of both mineral and synthetic Mn$_2$FeSbO$_6$ are listed in Table \ref{comp}.

\section{Conlusions}

The magnetic properties of both natural single-crystal mineral and synthetic ceramic Mn$_2$FeSbO$_6$ with ilmenite-type structure have been investigated. It was found that the natural mineral Mn$_2$FeSbO$_6$ undergoes a second-order magnetic phase transition at 268 K. The synthetic Mn$_2$FeSbO$_6$ orders at 264 K. Both samples order ferrimagneticly and show comparable values for the magnetic entropy change $\Delta S_M$ on the order of 2 J/kgK for 5 T and 0.5 J/kgK for 1 T field change, respectively. The adiabatic temperature change for the synthetic Mn$_2$FeSbO$_6$ near room-temperature under 1 T field change was estimate to  $\Delta T$ = 0.2 K. We believe that new functional magnetic materials with enhanced magnetocaloric properties based on Mn$_2$FeSbO$_6$ and its substitutions could be designed.

\section{Acknowledgments}
We thank the Swedish Energy Agency (STEM), Swedish Research Council (VR), the G\"oran Gustafsson Foundation, the Swedish Foundation for International Cooperation in Research and Higher Education (STINT), and the Russian Foundation for Basic Research for financial support. We are grateful to Dr. Henrik Skogby from the Swedish Museum of Natural History in Stockholm, Sweden, for providing melanostibite mineral.

\bibliographystyle{model1-num-names}


%

\end{document}